\documentclass[a4paper,aps,preprintnumbers,showpacs,superscriptaddress,nofootinbib,amsmath,amssymb,twocolumn]{revtex4}
\usepackage[dvips]{graphics}
\usepackage{cmap}
\usepackage[utf8]{inputenc}
\usepackage[T1]{fontenc}

\def\imo{i}

\def\K{{\cal K}}

\newcommand{\ie}{{i.e.,}~}

\begin{document}

\title{Quasinormal modes of renormalization group improved Dymnikova regular black holes}
\author{R. A. Konoplya}\email{roman.konoplya@gmail.com}
\affiliation{Research Centre for Theoretical Physics and Astrophysics, \\ Institute of Physics, Silesian University in Opava, \\ Bezručovo náměstí 13, CZ-74601 Opava, Czech Republic}
\author{Z. Stuchlík}\email{zdenek.stuchlik@physics.slu.cz}
\affiliation{Research Centre for Theoretical Physics and Astrophysics, \\ Institute of Physics, Silesian University in Opava, \\ Bezručovo náměstí 13, CZ-74601 Opava, Czech Republic}
\author{A. Zhidenko}\email{olexandr.zhydenko@ufabc.edu.br}
\affiliation{Centro de Matemática, Computação e Cognição (CMCC), Universidade Federal do ABC (UFABC), \\ Rua Abolição, CEP: 09210-180, Santo André, SP, Brazil}
\author{A. F. Zinhailo}\email{antonina.zinhailo@physics.slu.cz}
\affiliation{Research Centre for Theoretical Physics and Astrophysics, \\ Institute of Physics, Silesian University in Opava, \\ Bezručovo náměstí 13, CZ-74601 Opava, Czech Republic}

\begin{abstract}
We find accurate quasinormal frequencies of a quantum corrected black hole constructed in the renormalization group theory via the coordinate-independent iterative procedure, leading to the Dymnikova regular black hole. We show that while the fundamental mode is only slightly affected by the quantum correction, the overtones change at a much stronger rate. This outburst of overtones occurs because of the deformation of the geometry of the Schwarzschild black hole solely near the event horizon. For finding accurate values of overtones we developed a general procedure allowing one to use the Leaver method to metrics which, initially, are not expressed in terms of rational functions.
\end{abstract}
\maketitle

\section{Introduction}
The development of a model for a black hole that incorporates quantum corrections is crucial, not only as a specific solution within an unknown and internally consistent theory of quantum gravity, but also due to the core issue concerning the final stage of Hawking evaporation and the central singularities of classical solutions. One way to address this issue is through the application of the concept of asymptotically safe gravity \cite{Reuter:2019byg}.

Quantum field theory provides the essential foundation for this issue. A straightforward way to examine its behavior is through the improvement of the renormalization group \cite{Bonanno:2001hi,Bonanno:2002zb,Rubano:2004jq,Koch:2013owa,Pawlowski:2018swz,Ishibashi:2021kmf,Chen:2022xjk}. This method involves integrating only the beta function for the gravitational coupling, while ignoring the variations of other couplings. This leads to the effective Newton constant, which is dependent on the energy scale $k$.
The energy-dependent Newton constant is then applied to the classical black-hole solution to obtain the quantum-corrected lapse function, which characterizes a regular black hole \cite{Bonanno:2000ep,Bonanno:2006eu,Falls:2010he,Falls:2012nd,Torres:2014gta,Torres:2014pea,Bonanno:2017zen,Adeifeoba:2018ydh,Held:2019xde}. The concept of energy-dependent physical laws in the action is a common feature of quantum field theory.

In addition, the gravitational coupling relies on an arbitrary renormalization group scale denoted as $k$. Thus, a relationship between energy and radial coordinates must be established to express the resulting quantum-corrected black-hole metric. Currently, there are two options available to define the parameter $k$: The modified proper distance approach, resulting in the Bonanno-Reuter black-hole metric \cite{Bonanno:2000ep}, and the power of the Kretschmann scalar approach \cite{Pawlowski:2018swz,Adeifeoba:2018ydh}, resulting in the Hayward metric \cite{Held:2019xde}. The latter was initially suggested as a toy-model solution to the singularity problem in the collapse and evaporation scenario \cite{Hayward:2005gi}.

Alternatively, Kazakov and Solodukhin suggested a self-consistent method for deriving the quantum-corrected black-hole metric using the renormalization group approach \cite{Kazakov:1993ha}. There, neglecting nonspherical deformations of the background the problem was solved nonperturbatively. Quasinormal modes and gray-body factors were analyzed for this case in \cite{Konoplya:2019xmn,Saleh:2016pke}. However, the Kazakov-Solodukhin solution has not eliminated the singularity altogether, but instead relocated it to a finite distance from the center.

Quasinormal modes \cite{Konoplya:2011qq,Kokkotas:1999bd}, a fundamental property of black holes determined solely by their background parameters rather than perturbations, play a dominant role in the decay of perturbations at later stages and can be observed by gravitational interferometers~\cite{ligo1,ligo2,ligo3}. However, the wide range of possible angular momentum and mass values for observed black holes leaves ample room for modified gravity theories~\cite{Konoplya:2016pmh}. Consequently, a large body of literature exists on quasinormal modes for numerous regular black-hole models \cite{Bronnikov:2012ch,Lopez:2022uie,Lan:2022qbb,Rincon:2020cos,Jusufi:2020odz,Lan:2020fmn,Hendi:2020knv,MahdavianYekta:2019pol,Panotopoulos:2019qjk,Saleh:2018hba,Wu:2018xza,Li:2016oif,Fernando:2015fha,Toshmatov:2015wga,Li:2013fka,Lin:2013ofa,Flachi:2012nv}.
Extensive research on the spectrum of the Hayward solution has been carried out \cite{Flachi:2012nv,Li:2013fka,Lin:2013ofa,Toshmatov:2015wga,Roy:2022rhv}, and the quasinormal modes of the Bonanno-Reuter black hole were analyzed in \cite{Rincon:2020iwy,Panotopoulos:2020mii,Li:2013kkb}, along with an approximate truncated version in~\cite{Liu:2012ee}. However, many of the most critical and fascinating features of the quasinormal spectrum for both models were overlooked in all the aforementioned studies, in our view.

The most fascinating behavior occurs in the overtones, which were mostly neglected in the prior studies. While it is generally thought that the main contribution to the signal is from the fundamental mode, recent research in~\cite{Giesler:2019uxc} (and later examined in~\cite{Oshita:2021iyn,Forteza:2021wfq}) has shown that up to ten initial overtones are required to replicate the ringdown of accurate numerical relativity simulations at all stages, not just the final phase. This finding also indicates that the actual quasinormal ringing starts earlier than anticipated. The primary motivation for the renormalization group method employed here is obviously linked to quantum corrections, which are assumed to be quite small for astrophysical black holes. Nonetheless, our observation may have a wider interpretation, including sizable black holes, and the aspects of the overtones' behavior we discovered could be valuable. Additionally, we will argue that some of the traits found here should be universally applicable to a quantum corrected black hole.

The overtones for the above two choices of the identification has been recently studied in \cite{Konoplya:2022hll}. There it was shown that the quasinormal modes exhibit similar qualitative and even close quantitative characteristics for both types of the above identification for $k$. We have observed that the deviation of the fundamental mode from its Schwarzschild limit may be several times larger than previously reported in \cite{Rincon:2020iwy}. Notably, the overtones exhibit a striking deviation from the Schwarzschild limit, reaching as high as several hundred percent, even when the fundamental mode closely approximates the Schwarzschild limit. This result is due to the fact that both metrics are almost identical to the Schwarzschild metric everywhere except in a small region near the event horizon, which is crucial for the overtones. Additionally, the spectrum of both metrics includes nonoscillatory purely imaginary modes, which can emerge at the second overtone for certain parameter values \cite{Konoplya:2022hll}.

Here, instead of choosing this or that identification for $k$, we make use of the coordinate independent approach developed in~\cite{Platania:2019kyx}.
A solution obtained within this approach coincides with the Dymnikova-type black-hole spacetime \cite{Dymnikova:1992ux}, and it takes into account the backreaction effects of the running of the Newton coupling. The renormalization group approach, which is independent of coordinates, has been used to derive the spherically symmetric black-hole metric based on curvature invariants~\cite{Held:2021vwd}.

We will demonstrate that the overtones are again highly sensitive to the near-horizon asymptotic, which results in a substantial deviation of their real parts from their Schwarzschild limit, while the fundamental mode's deviation is typically extremely small. To put it differently, in the case where the geometry of the quantum-corrected black holes closely resembles the Schwarzschild geometry everywhere except a small region near the event horizon, the overtones significantly differ from their Schwarzschild counterparts. This observation seems to be linked to a similar phenomenon called the ``quasinormal modes' instability'', which was discovered in \cite{Jaramillo:2020tuu}. It was observed that while small perturbations to the initial linearized Einstein wave equations for the Schwarzschild background have little effect on the fundamental mode and the first few overtones, higher overtones undergo significant changes. In other words, highly damped modes are sensitive to slight variations in the wave equation. The work \cite{Jaramillo:2021tmt} suggests that this instability in the quasinormal modes might be detectable
 in the gravitational signal in the future.

We will perform a comprehensive analysis of the quasinormal ringing, including the behavior of overtones, for the regular black-hole solutions presented above, using test scalar, electromagnetic, and Dirac fields. While it is possible to examine the gravitational perturbations for these vacuum solutions of the Einstein equation, the primary assumption of the renormalization group approach is that the leading correction to the background metric results from the running gravitational coupling, which implies the existence of other correction terms that were ignored. Thus, there is no indication that small perturbations of the black-hole background spacetime will remain much smaller than the unknown terms that were initially neglected in the background. Furthermore, we know that gravitational perturbations are typically qualitatively similar to those for test fields and often coincide with the latter in the high-frequency (eikonal) regime.

In order to find quasinormal modes we will use the WKB formula, Bernstein spectral method, and Leaver method. However, only the latter allows one to find overtones with guaranteed accuracy. Therefore, our work has also a purely technical purpose: to develop a general procedure for finding accurate quasinormal modes by the Leaver method, when the metric coefficients initially are not represented by rational functions. This procedure can be further applied not only to analytical metrics like that describing the Dymnikova black hole \cite{Dymnikova:1992ux}, but also for spacetimes obtained numerically.

Our paper is organized as follows: In Sec.~\ref{sec:RGimprovement} we summarize the coordinate independent renormalization group approach developed in~\cite{Platania:2019kyx}, which leads to the Dymnikova black hole \cite{Dymnikova:1992ux}. Sec.~\ref{sec:parametrization} is devoted to representation of the metric function, in terms of rational functions, appropriate for further usage of the Leaver method. Sec.~\ref{sec:wavelike} briefly relates the properties of the wavelike equation, while in Sec.~\ref{sec:methods} we discuss the methods used for finding quasinormal modes'. In Sec.~\ref{sec:QNMs} we discuss the obtained numerical results for quasinormal frequencies and, finally, in the Conclusion we summarize the obtained results and mention open questions.

\section{Dymnikova black hole constructed via iterative RG-improvement procedure}\label{sec:RGimprovement}
The metric of the spherically symmetric spacetime has the following form:
\begin{equation}\label{metric}
ds^2 = -f(r) d t^2+ f^{-1}(r) dr^2 + r^2 (\sin^2 \theta d\phi^2+d\theta^2),
\end{equation}
where the metric for the Schwarzschild case is
\begin{equation}\label{lapse}
f(r)=1-\frac{2mG_0}{r} \;.
\end{equation}
We will begin with the usual Schwarzschild spacetime.
The form of the metric is described by the line element~\eqref{metric} and constitutes a solution to the Einstein vacuum equations with lapse function \eqref{lapse}:
\begin{equation}\label{startsys}
\begin{cases}
R_{\mu\nu}-\frac{1}{2}R\,g_{\mu\nu}=0, \\[0.2cm]
f_{(0)}(r)=1-\dfrac{2\,m\,G_{0}}{r}.
\end{cases}
\end{equation}
The perturbation of \eqref{startsys} via replacing the Newton's constant $G_0$ by its running analogue can be expressed as follows:
\begin{equation}\label{firstrepl}
G_{0}\to G(r)=\frac{G_0}{1+g_\ast^{-1} G_0 \, k^2(r)}\;,
\end{equation}
where $k(r)$ is a cutoff function such that $k(r)$ goes to zero as $r\to\infty$. The latter condition guarantees that the classical Schwarzschild metric is reproduced in the classical region $\bar{g}^{\mu\nu}\equiv\langle g^{\mu\nu}\rangle_k$ for $k\ll M_{Pl}$ and $r\gg l_{Pl}$. The replacement~\eqref{firstrepl} gives rise to a new metric of the form~\eqref{metric}, with the following lapse function
\begin{equation}
f(r)=1-\frac{2mG[k(r)]}{r}\;.
\end{equation}
The latter modified spacetime can be considered as an exact solution to the Einstein equations in the presence of the effective energy-momentum tensor,
\begin{equation}\label{Teff}
T_{\mu\nu}^{\text{eff}}=(\rho+p)(l_\mu n_\nu+l_\nu n_\mu)+p g_{\mu\nu}\;.
\end{equation}
The null vectors $l_\mu$ and $n_\mu$ obey the normalization condition $l_\mu n^\mu=-1$. The energy density $\rho$ and pressure $p$ are generated by the variation of the Newton's constant with the radial coordinate $r$,
\begin{equation}\label{qgfluid}
\rho=\frac{m\,G'(r)}{4\pi r^2\,G(r)}\;\;,\qquad p=-\frac{m\,G''(r)}{8\pi r\,G(r)}\;.
\end{equation}
The effective energy-momentum tensor~\eqref{Teff} can be interpreted as an effect from the vacuum polarization of quantum gravitational field. Then, the energy-density~$\rho$ should be considered as an effective quantum-gravitational self-energy. The quantum system is self-consistent: a small deviation of the Newton's constant induces successive back-reactions of the semiclassical background which, induces further deviations of the Newton's coupling. Following this way, the energy-density~$\rho\propto\partial_r \mathrm{log}\,G(r)$ should provide a kind of measure for quantum effects inside the horizon, and it can be applied to self-consistent construction of the cutoff function~$k_{(n+1)}(r)$ for $n>1$. The renormalization group improvement iteration is then formalized in the following way.

The first step of the iteration is the classical Schwarzschild metric, which is characterized by the parameters
\begin{equation}
k_{(0)}(r)=0 \quad\Rightarrow\quad G_{(0)}=G_0\;\;,\;\; T_{\mu\nu}^{\mathrm{eff}(0)}=0\;.
\end{equation}
The second step is defined by the replacement~\eqref{firstrepl}, where $k=k_1(r)$ is chosen arbitrarily and serves as an initial condition for the perturbation of the system \eqref{startsys}. Consequent steps of the iteration procedure, $n>1$, are defined by the following replacement:
\begin{equation}
G_{(n)}\to G_{(n+1)}(r)=\frac{G_0}{1+g_\ast^{-1} G_0 \, k_{(n+1)}^2(r)}\;\;,
\end{equation}
where the cutoff function $k_{(n+1)}(r)$ is constructed as a functional of the energy-density $\rho_{(n)}(r)$ generated by the variation of $G(r)$ in the previous step,
\begin{equation}\label{selfcutoff}
k_{(n+1)}^2(r)\equiv\mathcal{K}[\rho_{(n)}(r)]\;.
\end{equation}
The above procedure generates a sequence of the modified Einstein equations, admitting solutions of the form \eqref{metric} with the lapse function $f_{(n)}(r)$.

Following~\cite{Platania:2019kyx} in this way, it turns out that in the limit $n \rightarrow \infty$ the metric function coincides with the Dymnikova black hole \cite{Dymnikova:1992ux}
\begin{equation}\label{f(r)}
f(r) = 1-\frac{2 M}{r} \left(1-e^{-\dfrac{r^3}{2 l_{cr}^2 M}}\right).
\end{equation}
Here $l_{cr}$ is a critical length scale below which the modifications owing to the running of the Newton’s constant become negligible.
The maximal value of $l_{cr}$ allowing for existence of the event horizon is
$$l_{cr} \approx 1.138 M,$$
where $M\equiv G_0 m$ is the mass measured in units of length.

\begin{figure}
\resizebox{\linewidth}{!}{\includegraphics{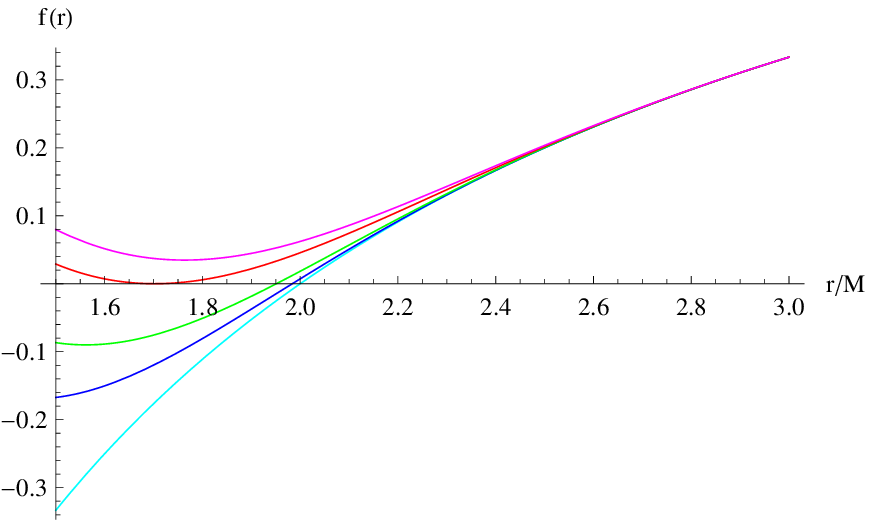}}\\
\resizebox{\linewidth}{!}{\includegraphics{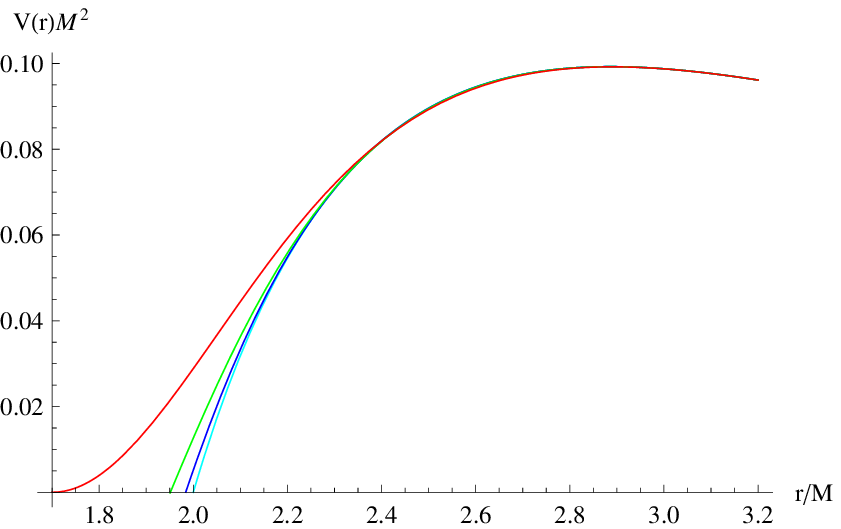}}
\caption{The metric function $f(r)$ (upper panel) and the effective potential for $\ell=1$ scalar field perturbations (lower panel): $l_{cr} =0.01M$ (bottom, cyan), $l_{cr} =0.9M$ (blue), $l_{cr} =M$ (green), $l_{cr} =1.138M$ (red), $l_{cr} =1.2M$ (magenta).}\label{fig:metric}
\end{figure}

An example of the metric functions and effective potentials are shown in Fig.~\ref{fig:metric}. When $l_{cr}\to0$, the Schwarzschild limit is reproduced.

\section{Parametrized metric as a rational function of the radial coordinate}\label{sec:parametrization}
In order to approximate our metric functions by rational functions of the radial coordinate we will use the parametrization approach \cite{Rezzolla:2014mua} for characterizing the spacetime of spherically symmetric black holes in generic metric theories of gravity. Unlike other methods, which employed a Taylor series expansion in powers of $M/r$, it utilizes continued-fraction expansions in terms of a compactified radial coordinate. This expansion provides superior convergence properties and allows us to approximate various metric theories using only a few coefficients. The parametrization was also extended to the axially-symmetric spacetimes \cite{Konoplya:2016jvv,Younsi:2016azx}, which means that the same approach is, at least in principle, can be used for rotating black holes.

Following \cite{Rezzolla:2014mua}, we introduce the following dimensionless variable:
\begin{equation}
x \equiv 1-\frac{r_0}{r},
\end{equation}
where $r_0$ is the event horizon radius, so that \mbox{$0\leq x\leq1$} spans the radial coordinate between the horizon and spatial infinity.

Then we approximate the metric function $f(r)$,
\begin{equation}\label{fapprox}
f\left(\frac{r_0}{1-x}\right)=x A(x),
\end{equation}
where $A(x)>0$ is defined as follows:
$$
A(x)=1-\epsilon (1-x)+(a_0-\epsilon)(1-x)^2+{\tilde A}(x)(1-x)^3\,,
$$
where the coefficient $\epsilon$ measures the deviation of $r_0$ from the Schwarzschild radius $2 M$,
\begin{equation}\label{epsilon}
\epsilon = \frac{2 M-r_0}{r_0},
\end{equation}
and the post-Newtonian parameter $a_0=0$.

The function ${\tilde A}$ is introduced through infinite continued fraction in order to describe the metric near the horizon (\ie for $x \simeq 0$),
\begin{equation}\label{Adef}
{\tilde A}(x)=\frac{a_1}{\displaystyle 1+\frac{\displaystyle a_2x}{\displaystyle 1+\frac{\displaystyle a_3x}{\displaystyle 1+\ldots}}},
\end{equation}
where $a_1, a_2, a_3, \ldots$ are dimensionless constants matching the near-horizon behavior of the metric function,
\begin{eqnarray}
a_1&=&\epsilon\left(2-\frac{3r_0^3}{2 l_{cr}^2 M}\right),\\\nonumber
a_2&=&\frac{9r_0^6 - 48 r_0^3 l_{cr}^2 M + 24 l_{cr}^4 M^2}{4 l_{cr}^2 M (3 r_0^3-4 l_{cr}^2 M)},\ldots,
\end{eqnarray}
and the closed form can be obtained for any $a_i$.

\begin{figure}
\resizebox{\linewidth}{!}{\includegraphics{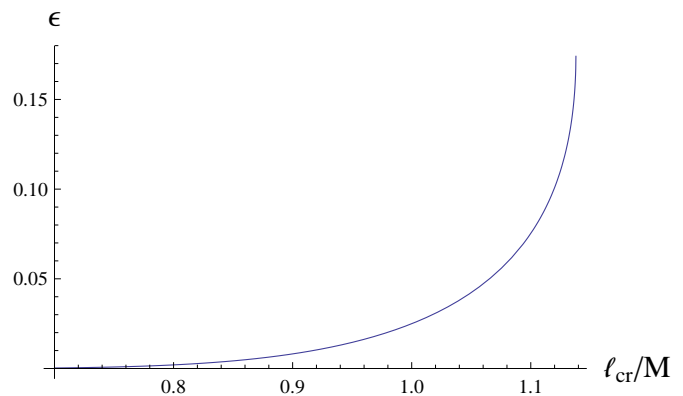}}
\caption{The value of $\epsilon$ as a function of $l_{cr}$.}\label{fig:epsilon}
\end{figure}
Since the horizon radius is a solution of a nonalgebraic equation $f(r_0)=0$, the value of $\epsilon$ defined in Eq.~(\ref{epsilon}) satisfies
\begin{equation}\label{epsilonequation}
  \epsilon=(\epsilon +1) e^{-\dfrac{4 M^2}{l_{cr}^2 (\epsilon +1)^3}}.
\end{equation}
From Eq.~(\ref{epsilonequation}) one can find $\epsilon$ numerically for any given $l_{cr}/M$ (see Fig.~\ref{fig:epsilon}).

In order to approximate the metric function we consider a finite number $n$ of terms in Eq.~(\ref{Adef}), i.e. we consider $a_{n+1}=0$. Then the metric function (\ref{fapprox}) is a rational function of $x$ and, therefore, of the radial coordinate $r$.
The Mathematica\textregistered{} notebook with the approximate metric is available as an ancillary file.

\section{The wavelike equations}\label{sec:wavelike}
The general relativistic equations for the scalar ($\Phi$), electromagnetic ($A_\mu$), and Dirac ($\Upsilon$) fields can be written in the following form:
\begin{subequations}\label{coveqs}
\begin{eqnarray}\label{KGg}
\frac{1}{\sqrt{-g}}\partial_\mu \left(\sqrt{-g}g^{\mu \nu}\partial_\nu\Phi\right)&=&0,
\\\label{EmagEq}
\frac{1}{\sqrt{-g}}\partial_{\mu} \left(F_{\rho\sigma}g^{\rho \nu}g^{\sigma \mu}\sqrt{-g}\right)&=&0\,,
\\\label{covdirac}
\gamma^{\alpha} \left( \frac{\partial}{\partial x^{\alpha}} - \Gamma_{\alpha} \right) \Upsilon&=&0,
\end{eqnarray}
\end{subequations}
where $F_{\mu\nu}=\partial_\mu A_\nu-\partial_\nu A_\mu$ is the electromagnetic tensor, $\gamma^{\alpha}$ are noncommutative gamma matrices and $\Gamma_{\alpha}$ are spin connections in the tetrad formalism.
After separation of the variables the above equations (\ref{coveqs}) take the Schrödinger wavelike form \cite{Kokkotas:1999bd,Berti:2009kk,Konoplya:2011qq}:
\begin{equation}\label{wave-equation}
\dfrac{d^2 \Psi}{dr_*^2}+(\omega^2-V(r))\Psi=0,
\end{equation}
where the ``tortoise coordinate'' $r_*$ is defined as follows:
\begin{equation}
dr_*\equiv\frac{dr}{f(r)}.
\end{equation}

The effective potentials for the scalar ($s=0$) and electromagnetic ($s=1$) fields have the form
\begin{equation}\label{potentialScalar}
V(r)=f(r) \frac{\ell(\ell+1)}{r^2}+\left(1-s\right)\cdot\frac{f(r)}{r}\frac{d f(r)}{dr},
\end{equation}
where $\ell=s, s+1, s+2, \ldots$ are the multipole numbers.
For the Dirac field ($s=1/2$) one has two isospectral potentials,
\begin{equation}
V_{\pm}(r) = W^2\pm\frac{dW}{dr_*}, \quad W\equiv \left(\ell+\frac{1}{2}\right)\frac{\sqrt{f(r)}}{r}.
\end{equation}
The isospectral wave functions can be transformed one into another by the Darboux transformation,
\begin{equation}\label{psi}
\Psi_{+}=q \left(W+\dfrac{d}{dr_*}\right) \Psi_{-}, \quad q=const,
\end{equation}
so that it is sufficient to calculate quasinormal modes for only one of the effective potentials. We will do that for $V_{+}(r)$ because the WKB method works better in this case.

\section{Methods used for calculations of quasinormal modes}\label{sec:methods}
Quasinormal modes $\omega_{n}$ of asymptotically flat black holes are proper oscillation frequencies corresponding to the solutions of the master wave equation (\ref{wave-equation}), when the purely outgoing waves at both infinities are imposed:
\begin{equation}
\Psi \propto e^{-\imo \omega t \pm \imo \omega r_*}, \quad r_* \to \pm \infty.
\end{equation}
Here we will briefly review the three methods which will be used for calculations of quasinormal frequencies: the WKB method, Bernstein spectral method, and Frobenius method.

\subsection{WKB method}
In the frequency domain we will use the semianalytic WKB approach applied by Will and Schutz \cite{Schutz:1985km} for finding quasinormal modes. The Will-Schutz formula was extended to higher orders in \cite{Iyer:1986np,Konoplya:2003ii,Matyjasek:2017psv} and made even more accurate when using the Padé approximants \cite{Matyjasek:2017psv,Hatsuda:2019eoj}.
The general WKB formula has the form \cite{Konoplya:2019hlu}
\begin{eqnarray}
&&\omega^2=V_0+A_2(\K^2)+A_4(\K^2)+A_6(\K^2)+\ldots \\\nonumber
&&-\imo\K\sqrt{-2V_2}\left(1+A_3(\K^2)+A_5(\K^2)+A_7(\K^2)+\ldots\right),
\end{eqnarray}
where $\K=n+1/2$ is half-integer. The corrections $A_k(\K^2)$ of the order $k$ to the eikonal formula are polynomials of $\K^2$ with rational coefficients and depend on the values of higher derivatives of the potential $V(r)$ in its maximum. In order to increase the accuracy of the WKB formula, we will follow the procedure of Matyjasek and Opala \cite{Matyjasek:2017psv} and use the Padé approximants. Here we will use the sixth-order WKB method with $\tilde{m}=4$, where $\tilde{m}$ is defined in \cite{Matyjasek:2017psv,Konoplya:2019hlu}, because this choice provides the best accuracy in the Schwarzschild limit, and there is hope that this will be the case for more general metrics.

\subsection{Frobenius method}
In order to find accurate values of quasinormal modes we use the method proposed by Leaver~\cite{Leaver:1985ax}. The wavelike equation~(\ref{wave-equation}) always has a regular singularity at the event horizon $r=r_0$ and the irregular singularity at spatial infinity $r=\infty$. We introduce the new function,
\begin{equation}\label{reg}
\Psi(r)=e^{\imo\omega r}r^{2M\imo\omega}\left(1-\frac{r_0}{r}\right)^{-\imo\omega/f'(r_0)}y(r),
\end{equation}
so that $y(r)$ is regular for $r_0\leq r<\infty$, once $\Psi(r)$ corresponds to the purely outgoing wave at spatial infinity and the purely ingoing wave at the event horizon. Therefore, we are able to represent $y(r)$ in terms of the Frobenius series:
\begin{equation}\label{Frobenius}
y(r)=\sum_{k=0}^{\infty}a_k\left(1-\frac{r_0}{r}\right)^k.
\end{equation}

Using Eq.~(\ref{wave-equation}), all the coefficients $a_{k>0}$ can be expressed in terms of $a_0$. Unfortunately, since the function $f(r)$ in (\ref{f(r)}) is not a rational function of $r$, one cannot derive a finite recurrence relation. However, once we use the approximation using the finite continued fraction (\ref{fapprox}), we can find a finite recurrence relation, which can be reduced to the one of three terms via the Gaussian elimination (see, for example, \cite{Konoplya:2011qq} for a detailed description of the procedure). The number of terms in the recurrence relation increases with the approximation order $n$ in Eq.~(\ref{Adef}). For the Frobenius series defined in~(\ref{reg}), the number of terms in the recurrence relation is $10+n$. This number could be further decreased by taking into account the inner horizon radius in the Frobenius series. However, the inner horizon for the approximated metric differs from the one for the accurate metric and depends on the approximation order. Therefore, in order to have a general procedure, which can be applied for any approximation order, we have employed the simplest Frobenius series~(\ref{reg}), which depends on the horizon radius only.

Then, using the recurrence relation coefficients, we find the equation with the infinite continued fraction with respect to $\omega$, which is satisfied iff the series (\ref{Frobenius}) converges at $r=\infty$, i.e., when $\Psi(r)$ obeys the quasinormal boundary conditions. In order to calculate the infinite continued fraction we also use the Nollert improvement~\cite{Nollert:1993zz}, which was generalized in~\cite{Zhidenko:2006rs} for an arbitrary number of terms in the recurrence relation. Namely, for each given $\omega$ we calculate numerically the values of $C_0(\omega)$, $C_1(\omega)$, and $C_2(\omega)$ in Eq.~(17) of~\cite{Zhidenko:2006rs} and use the approximation for the value of the infinite continued fraction tail, making sure that the result does not change if we increase the position of the tail.

\begin{figure*}
\resizebox{\linewidth}{!}{\includegraphics{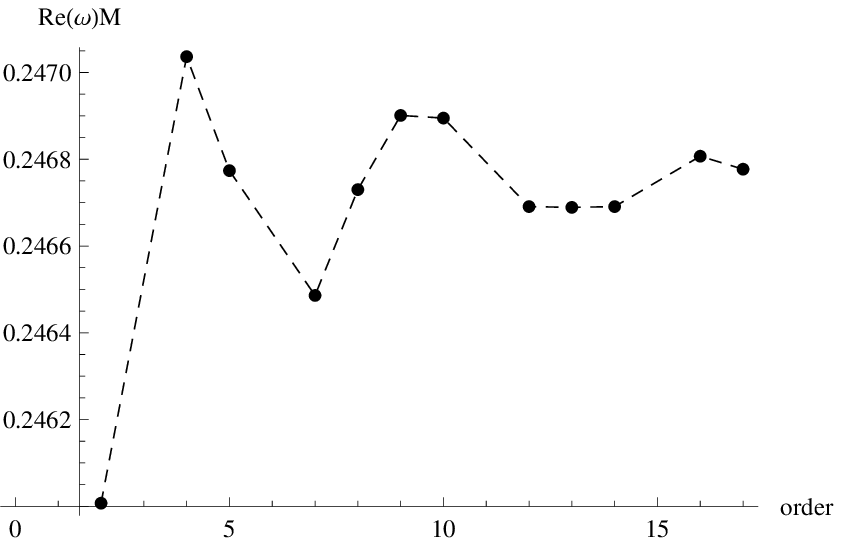}\includegraphics{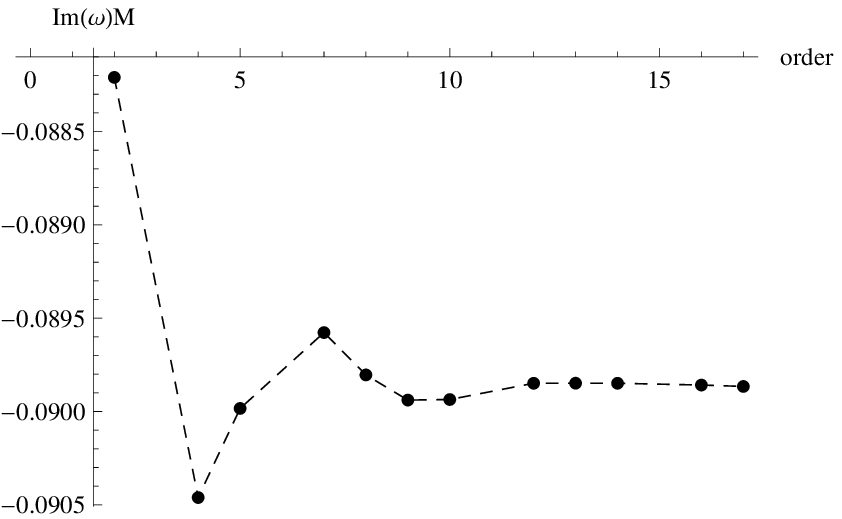}}
\caption{Convergence of the dominant quasinormal mode for $\ell=1$ electromagnetic perturbations with respect to the approximation order ($l_{cr}=M=1$).}\label{fig:convergence}
\end{figure*}

For nonsmall values of $l_{cr}$ the behavior of the metric function near the event horizon differs significantly from the Schwarzschild black hole (see Fig.~\ref{fig:metric}). The function $f(r)$ grows slowly, and, when approximating the function by a rational function of $r$ the resulting geometry implies that the inner horizon appears even though the accurate metric does not have one. This leads to an additional singular point in the wavelike equation~(\ref{wave-equation}), which diminishes the convergence radius of the series~(\ref{Frobenius}). It is possible to avoid this problem by modifying the Frobenius series as described in~\cite{Konoplya:2007jv}.
Instead, since generally the additional singular points depend not only on the value of $l_{cr}$ but also on the order of approximation (\ref{Adef}), we continue the series (\ref{Frobenius}) through some
positive real midpoints as described in~\cite{Rostworowski:2006bp}. Unfortunately, such a complication of the singular point structure seems to result in the slower convergence of the values of the quasinormal modes, with respect to the approximation order (see Fig.~\ref{fig:convergence}).

\subsection{Bernstein polynomial spectral method}
Following \cite{Fortuna:2020obg}, we introduce the compact coordinate $u$,
$$u\equiv\frac{1}{r},$$
and represent $y(u)$ as a sum,
\begin{equation}\label{Bernsteinsum}
y(u)=\sum_{k=0}^NC_kB_k^N(u),
\end{equation}
where
$$B_k^N(u)\equiv\frac{N!}{k!(N-k)!}u^k(1-u)^{N-k}$$
are the Bernstein polynomials.

Substituting (\ref{reg}) into (\ref{wave-equation}) and using a Chebyschev collocation grid of $N+1$ points, we obtain a set of linear equations with respect to $C_k$, which has nontrivial solutions iff the corresponding coefficient matrix is singular. The problem is reduced to the eigenvalue problem of a matrix pencil with respect to $\omega$, which can be solved numerically. Once the eigenvalue problem is solved, one can calculate the corresponding coefficients $C_k$ and explicitly determine the polynomial (\ref{Bernsteinsum}), which approximates the solution to the wave equation \cite{Fortuna:2020obg}.

In order to exclude the spurious eigenvalues, which appear due to finiteness of the polynomial basis in (\ref{Bernsteinsum}), we compare both the eigenfrequencies and corresponding approximating polynomials for different values of $N$. Namely, for the coinciding eigenfrequencies, $\omega^{(1)}$ and $\omega^{(2)}$, obtained, respectively, for $N=N^{(1)}$ and $N=N^{(2)}$, we calculate
$$1-\frac{|\langle y^{(1)}\;|\;y^{(2)} \rangle|^2}{||y^{(1)}||^2||y^{(2)}||^2}=\sin^2\alpha,$$
where $\alpha$ is the angle between the vectors $y^{(1)}$ and $y^{(2)}$ in the $L^2$-space. If $\alpha$ is sufficiently small, then the difference between $\omega^{(1)}$ and $\omega^{(2)}$ provides the error estimation\footnote{The Wolfram Mathematica\textregistered{} package with the implementation of the Bernstein spectral method \cite{Konoplya:2022zav} is publicly available from \url{https://arxiv.org/src/2211.02997/anc}.}.

\begin{table*}
\begin{tabular}{|l||c|c||c|c||c|}
 \hline
  $l_{cr}$ & WKB & error & Bernstein & error & Frobenius \\
 \hline
  $0.5$  & $0.110892-0.104608\imo$ & $0.34230\%$ & $0.110558-0.105125\imo$ & $0.16111\%$ & $0.11046-0.10490\imo$ \\
  $0.9$  & $0.098211-0.093779\imo$ & $9.92174\%$ & $0.110071-0.103336\imo$ & $0.20876\%$ & $0.10994-0.10305\imo$ \\
  $1.0$  & $0.096781-0.092553\imo$ & $9.66451\%$ & $0.108911-0.100335\imo$ & $0.13202\%$ & $0.10890-0.10014\imo$ \\
  $1.1$  & $0.095144-0.090187\imo$ & $7.77728\%$ & $0.103852-0.096527\imo$ & $0.21895\%$ & $0.10416-0.09657\imo$ \\
  \hline
\end{tabular}
\caption{The fundamental ($n=0$) quasinormal mode for $\ell=0$, scalar field perturbations; the relative error is calculated assuming that the Frobenius values are accurate.}
\end{table*}

\begin{table}
\begin{tabular}{|l|c|c|}
 \hline
  $l_{cr}$ & WKB & Bernstein \\
 \hline
  $0.5$  & $0.182749-0.097081\imo$ & $0.190340-0.097486\imo$  \\
  $0.9$  & $0.174997-0.097650\imo$ & $0.187575-0.094316\imo$  \\
  $1.0$  & $0.173153-0.096513\imo$ & $0.182531-0.089763\imo$  \\
  $1.1$  & $0.167839-0.091445\imo$ & $0.167347-0.091296\imo$  \\
  \hline
\end{tabular}
\caption{The fundamental ($n=0$) quasinormal mode for $\ell=1/2$, Dirac field perturbations; $M=1$.}
\end{table}

\begin{table*}
\begin{tabular}{|l||c|c||c|c||c|}
 \hline
  $l_{cr}$ & WKB & error & Bernstein & error & Frobenius \\
 \hline
  $0.5$ & $0.29293 - 0.09766\imo$ & $0.00324\%$ & $0.29294 - 0.09767 \imo$ & $0.00324\%$ & $0.29294-0.09766\imo$ \\
  $0.9$ & $0.29125 - 0.09899\imo$ & $0.61139\%$ & $0.29204 - 0.09725 \imo$ & $0.01657\%$ & $0.29199-0.09726\imo$ \\
  $1.0$ & $0.28995 - 0.09811\imo$ & $0.64575\%$ & $0.29074 - 0.09631 \imo$ & $0.00980\%$ & $0.29077-0.09631\imo$ \\
  $1.1$ & $0.28782 - 0.09616\imo$ & $0.57349\%$ & $0.28854 - 0.09470 \imo$ & $0.06615\%$ & $0.28874-0.09468\imo$ \\
  \hline
\end{tabular}
\caption{The fundamental ($n=0$) quasinormal mode for $\ell=1$, scalar field perturbations; $M=1$; the relative error is calculated assuming that the Frobenius values are accurate.}
\end{table*}

\begin{table*}
\begin{tabular}{|l||c|c||c|c||c|}
 \hline
  $l_{cr}$ & WKB & error & Bernstein & error & Frobenius \\
 \hline
  0.5 & $0.248255 - 0.092497\imo$ & $0.00325\%$ & $0.248255-0.092492\imo$ & $0.00203\%$ & $0.24826-0.09249\imo$ \\
  0.9 & $0.246233 - 0.093069\imo$ & $0.77832\%$ & $0.247604-0.091552\imo$ & $0.02563\%$ & $0.24766-0.09159\imo$ \\
  1.0 & $0.244931 - 0.090477\imo$ & $0.74937\%$ & $0.246753-0.089864\imo$ & $0.01796\%$ & $0.2468~-0.08986\imo$ \\
  1.1 & $0.243683 - 0.088349\imo$ & $0.77155\%$ & $0.245052-0.087023\imo$ & $0.73291\%$ & $0.24516-0.08699\imo$ \\
  \hline
\end{tabular}
\caption{The fundamental ($n=0$) quasinormal mode for $\ell=1$, electromagnetic perturbations; $M=1$; the relative error is calculated assuming that the Frobenius values are accurate.}
\end{table*}

\section{Quasinormal modes}\label{sec:QNMs}
The crucial aspect for finding quasinormal modes not for the exact numerical or analytical solution, but for its parametrized approximation, is the convergence in orders of the parametrization. In Fig.~\ref{fig:convergence} we can see that in order to find the fundamental mode with sufficiently high accuracy, the truncation at about the 17th order of the parametrization is necessary. In that case the five decimal places are guaranteed for the fundamental mode, and, normally, a few decimal places for the first several overtones.
Unlike the Leaver method, the WKB and Bernstein polynomial methods can be applied directly to the exact exponential form of the metric, so that the error due truncation of the parametrization is excluded from their data. However, the WKB and Bernstein polynomial methods are not on equal footing here because the WKB formula converges only asymptotically, and its accuracy is, strictly speaking, unknown. That is why the Leaver and Bernstein polynomial methods are in very good concordance for the fundamental mode, as can be seen in tables I-IV, but the difference with WKB increases as the $l_{cr}$ approaches the extreme value. At the same time the Bernstein polynomial method is converging slowly when the overtone number $n$ is increased, so that the Leaver method remains the only efficient and quickly convergent tool for finding the overtones.

The fundamental mode is characterised by smaller oscillation frequency and damping rate, once the quantum correction represented by the parameter $l_{cr}$ is turned on. However, this is not so always for overtones which may have nonmonotonic dependence on $l_{cr}$ near the extremal limit.

\begin{table*}
\begin{tabular}{|l|c|c|c|c|}
  \hline
  $l_{cr}$            & $n=0$                 & $n=1$               & $n=2$             & $n=3$ \\
  \hline
  $0$  & $0.11046-0.10490\imo$ & $0.0861-0.3481\imo$ & $0.076-0.601\imo$ & $0.070-0.854\imo$ \\
  $0.7$               & $0.11042-0.10482\imo$ & $0.0846-0.3477\imo$ & $0.069-0.601\imo$ & $0.050-0.853\imo$ \\
  $0.75$              & $0.11039-0.10470\imo$ & $0.0827-0.3472\imo$ & $0.061-0.601\imo$ & $0.016-0.851\imo$ \\
  $0.8$               & $0.11034-0.10445\imo$ & $0.0792-0.3462\imo$ & $0.043-0.601\imo$ & $0.031-0.969\imo$ \\
  $0.85$              & $0.11016-0.10388\imo$ & $0.0732-0.3420\imo$ & $0.011-0.576\imo$ & $0.054-0.915\imo$ \\
  $0.9$               & $0.10994-0.10305\imo$ & $0.064~-0.337~\imo$ & $0.033-0.660\imo$ & $0.07~-0.87~\imo$ \\
  $0.95$              & $0.10959-0.10184\imo$ & $0.047~-0.333~\imo$ & $0.04~-0.64~\imo$ & $0.04~-0.84~\imo$ \\
  $1.0$               & $0.10890-0.10014\imo$ & $0.031~-0.376~\imo$ & $0.01~-0.60~\imo$ & $0.07~-1.03~\imo$ \\
  $1.05$              & $0.10737-0.09797\imo$ & $0.046~-0.362~\imo$ & $0.01~-0.69~\imo$ & $0.09~-1.06~\imo$ \\
  $1.1$               & $0.10416-0.09657\imo$ & $0.0356-0.3674\imo$ & $0.053-0.724\imo$ & $0.127-1.043\imo$ \\
   \hline
\end{tabular}
\caption{Quasinormal modes found by the Leaver method for $\ell=0$, scalar perturbations; $M=1$. The metric is approximated by the 17th-order parametrization. The Schwarzschild limit corresponds to $l_{cr} =0$.}\label{tabl:Frobenius:s=l=0}
\end{table*}

\begin{table*}
\begin{tabular}{|l|c|c|c|c|}
  \hline
  $l_{cr}$            & $n=0$                 & $n=1$               & $n=2$               & $n=3$ \\
  \hline
  $0$ & $0.29294-0.09766\imo$ & $0.2644-0.3063\imo$ & $0.2295-0.5401\imo$ & $0.203-0.788\imo$ \\
  $0.7$               & $0.29290-0.09766\imo$ & $0.2640-0.3063\imo$ & $0.2275-0.5405\imo$ & $0.197-0.790\imo$ \\
  $0.75$              & $0.29283-0.09765\imo$ & $0.2634-0.3063\imo$ & $0.2247-0.5409\imo$ & $0.188-0.792\imo$ \\
  $0.8$               & $0.29271-0.09760\imo$ & $0.2623-0.3061\imo$ & $0.2199-0.5410\imo$ & $0.173-0.794\imo$ \\
  $0.85$              & $0.29246-0.09748\imo$ & $0.2604-0.3054\imo$ & $0.212~-0.540~\imo$ & $0.155-0.786\imo$ \\
  $0.9$               & $0.29199-0.09726\imo$ & $0.2571-0.3037\imo$ & $0.200~-0.533~\imo$ & $0.14~-0.80~\imo$ \\
  $0.95$              & $0.29146-0.09688\imo$ & $0.2533-0.3020\imo$ & $0.184~-0.534~\imo$ & $0.14~-0.81~\imo$ \\
  $1.0$               & $0.29077-0.09631\imo$ & $0.2485-0.3002\imo$ & $0.170~-0.544~\imo$ & $0.09~-0.80~\imo$ \\
  $1.05$              & $0.28986-0.09556\imo$ & $0.2430-0.2990\imo$ & $0.157~-0.551~\imo$ & $0.11~-0.97~\imo$ \\
  $1.1$               & $0.28874-0.09468\imo$ & $0.2378-0.2981\imo$ & $0.144~-0.568~\imo$ & $0.159-0.964\imo$ \\
   \hline
\end{tabular}
\caption{Quasinormal modes found by the Leaver method for $\ell=1$, scalar perturbations; $M=1$. The metric is approximated by the 17th-order parametrization. The Schwarzschild limit corresponds to $l_{cr} =0$.}\label{tabl:Frobenius:s=0}
\end{table*}

\begin{table*}
\begin{tabular}{|l|c|c|c|c|}
  \hline
  $l_{cr}$            & $n=0$                 & $n=1$               & $n=2$             & $n=3$ \\
  \hline
  $0$ & $0.24826-0.09249\imo$ & $0.2145-0.2937\imo$ & $0.175-0.525\imo$ & $0.146-0.772\imo$ \\ 
  $0.7$               & $0.24823-0.09247\imo$ & $0.2141-0.2935\imo$ & $0.173-0.525\imo$ & $0.140-0.772\imo$ \\ 
  $0.75$              & $0.24814-0.09239\imo$ & $0.2135-0.2930\imo$ & $0.170-0.523\imo$ & $0.135-0.768\imo$ \\ 
  $0.8$               & $0.24804-0.09226\imo$ & $0.2125-0.2922\imo$ & $0.166-0.521\imo$ & $0.123-0.765\imo$ \\ 
  $0.85$              & $0.24790-0.09200\imo$ & $0.2112-0.2909\imo$ & $0.160-0.518\imo$ & $0.103-0.762\imo$ \\ 
  $0.9$               & $0.24766-0.09159\imo$ & $0.2088-0.2886\imo$ & $0.149-0.512\imo$ & $0.06~-0.74~\imo$ \\ 
  $0.95$              & $0.2473~-0.09085\imo$ & $0.206~-0.2849\imo$ & $0.13~-0.50~\imo$ & $0.06~-0.81~\imo$ \\ 
  $1.0$               & $0.2468~-0.08986\imo$ & $0.201~-0.280~\imo$ & $0.107-0.508\imo$ & $0.05~-0.8~~\imo$ \\ 
  $1.05$              & $0.24614-0.08855\imo$ & $0.195~-0.276~\imo$ & $0.107-0.514\imo$ & $0.04~-0.8~~\imo$ \\ 
  $1.1$               & $0.24516-0.08699\imo$ & $0.1892-0.2731\imo$ & $0.091-0.519\imo$ & $0.05~-0.87~\imo$ \\ 
   \hline
\end{tabular}
\caption{Quasinormal modes found by the Leaver method for $\ell=1$, electromagnetic perturbations; $M=1$. The metric is approximated by the 17th-order parametrization. The Schwarzschild limit corresponds to $l_{cr} =0$.}\label{tabl:Frobenius:s=1}
\end{table*}

From table~\ref{tabl:Frobenius:s=l=0} we can see that while the fundamental mode is changed by only a few percent, when $l_{cr}$ is changed from its Schwarzschild limit to the extremal one, the overtones can change significantly: already for the first overtone the oscillation frequency changes within tens of percent, while for the higher overtones, it can approach zero. For larger multipole ($\ell=1$) the effect takes place at higher overtones (see table~\ref{tabl:Frobenius:s=0}): the real part for $n=2$ changes within tens of percent and for $n=3$ can be more than two times smaller compared to the Schwarzschild one. For the electromagnetic perturbations presented in table~\ref{tabl:Frobenius:s=1} the third overtone is almost three times smaller in the near extremal regime than in the Schwarzschild limit. The effect is due to the deviation of the spacetime geometry solely near the event horizon \cite{Konoplya:2022hll,Konoplya:2022pbc,Konoplya:2022iyn}. The effective potential is very close to the Schwarzschild one near its peak and the fundamental mode, unlike the overtones, is usually determined by the scattering near the peak, being insensitive to the near-horizon deformations.

In Table~\ref{tabl:Frobenius:s=1} we present first four quasinormal modes of the Dymnikova black hole, calculated by the Frobenius method for the metric approximated with the 17th-order parametrization. The number of obtained decimal places becomes smaller for higher overtones and near-extreme values of $l_{cr}$. In this regime the metric function changes strongly near the event horizon but is very close to the Schwarzschild metric at a distance from it. Such type of metrics were called {\it nonmoderate} in \cite{Konoplya:2020hyk}; it requires many orders of the parametrization to be approximated sufficiently well. We conclude that the values of the quasinormal modes converge when increasing the approximation order, even for such nonmoderate black holes (see Fig.~\ref{fig:convergence}). Although the accuracy of our numerical approach becomes worse in this regime, the effect, \ie deviation from the Schwarzschild values, is clearly larger than the error of the approximation, so that we can claim that, while the variation of the dominant modes is insignificant, the higher overtones' frequencies change at a much stronger rate, as $l_{cr}$ approaches the extreme value.

\section{Conclusion}
In the present paper we have studied two issues. The first one is methodological: How to find accurate quasinormal modes, including overtones, for metric functions which, initially, are not expressed in terms of rational functions. The solution consists in approximation of the metric with the help of the continued fraction parametrization suggested in \cite{Rezzolla:2014mua}. One difficulty occurs in this way, which is related to accurate calculation of overtones: the higher overtone one needs to find, the more order of the parametrization is necessary to use. We further applied this procedure to find accurate frequencies both for the fundamental mode and several first overtones of the black hole constructed via the iterative coordinate independent procedure in the renormalization group approach~\cite{Platania:2019kyx}. We have found that while the fundamental mode changes by only a few percent, the overtones are highly sensitive to deformations of the geometry in the near-horizon zone induced by the quantum correction.

A similar approach of the parametrization and consequent usage of the Leaver method can be applied to axially-symmetric spacetimes, the general formalism for which was developed and studied in \cite{Konoplya:2016jvv,Younsi:2016azx}. Thus, the finding of accurate quasinormal frequencies for black holes, whose metric are given numerically or in terms of the nonrational functions, can be done according to the same principles for rotating black holes as well. Once a reliable axially-symmetric black-hole solution in the renormalization group theory is obtained, usage of the continued fraction expansion in the radial direction and expansion near the equatorial plane in the azimuthal one \cite{Konoplya:2016jvv,Younsi:2016azx} should allow the finding of accurate quasinormal frequencies then. However, such a task would be much more involved because of the problem of convergence in the angular sector.

\acknowledgments
A.~Z. was supported by Conselho Nacional de Desenvolvimento Científico e Tecnológico (CNPq).

\bibliographystyle{unsrt}
\bibliography{Bibliography}

\end{document}